\documentclass{article}
\usepackage{spconf,amsmath,graphicx}
\makeatletter
\renewcommand\section{\@startsection {section}{1}{\z@}%
	{-2.0ex \@plus -0.75ex \@minus -.2ex}%
	{1.ex \@plus.2ex \@minus -.2ex}%
	{\centering\bfseries}}
\renewcommand\subsection{\@startsection {subsection}{2}{\z@}%
	{-1.5ex \@plus -0.5ex \@minus -0.5ex}%
	{1.5ex \@plus.2ex}%
	{\bfseries}}
\renewcommand\subsubsection{\@startsection {subsubsection}{3}{\z@}%
	{-0.8ex \@plus -.2ex \@minus -.2ex}%
	{0.8ex \@plus.2ex \@minus -.2ex}%
	{}}
\makeatother
\usepackage{float}
\usepackage{amssymb}
\usepackage{booktabs}
\usepackage{enumitem}
\usepackage{pdfpages}
\usepackage[justification=centering]{caption}
\usepackage{subcaption}
\usepackage{mathtools}
\usepackage[export]{adjustbox}
\usepackage{hyperref}
\hypersetup{colorlinks, linkcolor=red}
\usepackage[absolute,showboxes]{textpos}

\TPMargin{5pt}

\hypersetup{
	colorlinks,
	citecolor=blue,
	linkcolor=red,
	urlcolor=red,
    citebordercolor=blue,
    filebordercolor=red,
    linkbordercolor=red
}

\newcommand{\copyrightstatement}{
	\begin{textblock}{0.95}(0.025,0.93)    
		\noindent
		\footnotesize
		\copyright \hspace*{0.5mm}Copyright 2020 IEEE. To be published in the IEEE 2020 International Conference on Acoustics, Speech, and Signal Processing (ICASSP 2020), scheduled for 4-9 May, 2020, in Barcelona, Spain. Personal use of this material is permitted. However, permission to reprint/republish this material for advertising or promotional purposes or for creating new collective works for resale or redistribution to servers or lists, or to reuse any copyrighted component of this work in other works, must be obtained from the IEEE. Contact: Manager, Copyrights and Permissions / IEEE Service Center / 445 Hoes Lane / P.O. Box 1331 / Piscataway, NJ 08855-1331, USA. Telephone: + Intl. 908-562-3966. 
	\end{textblock}
}
\title{LE\lowercase{t}-SNE: A Hybrid Approach To Data Embedding and Visualization of Hyperspectral Imagery}
\name{Megh Shukla$^{\star \dagger}$ \qquad Biplab Banerjee$^{\dagger\ddagger}$ \qquad Krishna Mohan Buddhiraju$^{\dagger}$}
\address{$^{\star}$Mercedes-Benz Research and Development India Pvt. Ltd.\\ 
	$^{\dagger}$Centre of Studies in Resources Engineering, Indian Institute of Technology Bombay
	\thanks{\hspace*{-0.5cm}${}^\ddagger$B. Banerjee was partially supported by SERB, DST (ECR/2017/000365)}}
\begin{document}
\copyrightstatement
\maketitle
\setlength{\belowdisplayskip}{2pt} \setlength{\belowdisplayshortskip}{0pt}
\setlength{\abovedisplayskip}{2pt} \setlength{\abovedisplayshortskip}{0pt}
\begin{abstract}
Hyperspectral Imagery (and Remote Sensing in general) captured from UAVs or satellites are highly voluminous in nature due to the large spatial extent and wavelengths captured by them. Since analyzing these images requires a huge amount of computational time and power, various dimensionality reduction techniques have been used for feature reduction. Some popular techniques among these falter when applied to Hyperspectral Imagery due to the famed curse of dimensionality. In this paper, we propose a novel approach, \textbf{LEt-SNE}, which combines graph based algorithms like t-SNE and Laplacian Eigenmaps into a model parameterized by a shallow feed forward network. We introduce a new term, \textit{Compression Factor,} that enables our method to combat the curse of dimensionality. The proposed algorithm is suitable for manifold visualization and sample clustering with labelled or unlabelled data. We demonstrate that our method is competitive with current state-of-the-art methods on hyperspectral remote sensing datasets in public domain.
\end{abstract}
\section{Introduction}
\label{sec:intro}
With the increasing availability of hyperspectral imagery, researchers face a challenging task of analyzing this data. Storing and processing this vast amount of data is cumbersome and expensive which leads us to an extensively studied topic, Dimensionality Reduction. The principle behind dimensionality reduction is the utilization of statistical information within the data to come up with a condensed representation for the same. A subset of dimensionality reduction, \textit{Manifold Learning} \cite{Cayton}, deals with the non-linear embedding of data in lower dimensions. When dealing with a high dimensional dataset such as hyperspectral imaging, we often encounter a phenomenon commonly known as the \textit{curse of dimensionality}; which entails that as the dimensionality $d$ of the dataset increases, the concept of \textit{neighbourhood} is lost. The distance between the farthest and the nearest samples is negligible when compared to the distance from a fixed sample to its nearest neighbor. This phenomenon is rigorously studied in \cite{Aggarwal, beyer, Francois, Hinneburg:2000}, with algorithms based on the euclidean distance suffering from the same. The focus of this paper is to create an algorithm that solves a three-fold problem: Manifold visualization, supervised clustering, and present a proof of concept for unsupervised clustering using image segmentation techniques. The core algorithm fuses a modification of t-SNE with Laplacian Eigenmaps into a model parameterized with a shallow fully connected neural network yielding quick encodings on unseen samples. To circumvent the curse of dimensionality, we introduce \textit{Compression factor}, which creates an illusion of modifying inter-sample distance. We compare our approach with state-of-the-art techniques on three open source remote sensing datasets: \textit{Indian Pines, Pavia University,} and \textit{Salinas} and present the results.\par
\textbf{Related Work: }Dimensionality reduction can be subdivided into broadly two categories. Feature selection algorithms such as Genetic Algorithms \cite{GA} and Ant Colony Optimization \cite{ACO} have been widely used, but do not provide information about the underlying manifold. Feature extraction algorithms such as PCA \cite{PCAapp} and LDA do not model non-linearities in the data, whereas non-linear methods such as kernel-PCA suffer from prohibitive time complexity \cite{Zhang}. They also capture the global structure of data at the cost of local variations in the manifold. Another limitation of LDA is that the dimensionality of the embeddings is bounded by the number of classes present in the dataset. Graph based algorithms such as \textit{Laplacian Eigenmaps} \cite{BelkinNiyogi} and \textit{Locally Linear Embedding (LLE)} \cite{Roweis2323} do not scale well with addition of new samples as they need to recompute the eigendecomposition to obtain new embeddings. \textit{t-SNE} \cite{vanDerMaaten2008}, though otherwise a beautiful visualization technique, fails to effectively deal with the curse of dimensionality. A less common variant of t-SNE is the parameteric t-SNE \cite{tSNEparam}, which uses Restricted Boltzmann Machines and pretraining which leads to a complicated training procedure. Recent approaches include \textit{UMAP} \cite{McInnes2018UMAPUM} which relies on projecting points along a Reimannian manifold, and Autoencoders \cite{51919, Hinton504, Vincent2010}. \textit{Spherical Stochastic Neighbor Encoding} \cite{sSNE}, constrains samples onto the surface of a  $\mathbb{R}^m$ unit hypersphere in a $\mathbb{R}^{m+1}$ space resulting in an ineffective use of the hyperspace. It remains to be seen if the sSNE can scale to large remote sensing datasets.
\section{Methodology}
\label{sec:method}
The key points considered when designing LEt-SNE include parameterization, computational time, the curse of dimensionality and the quality of embeddings produced. \par
Approaches using eigendecomposition as their solution need to recompute embeddings afresh when exposed to unseen samples. Parameterized models therefore have an advantage when mapping unseen samples since they model a function $Y=f(X, w)$, which takes as input $X$ with parameters $w$ to quickly compute the embeddings $Y$. A natural candidate for implementing this is a fully connected neural network architecture. The advantages presented by a neural network are multifold: they learn the task of feature extraction; stochasticity in the training process with the use of mini-batch weight updates acts as a regularizing mechanism. Mini-batches lead to faster convergence as the weights are periodically updated without waiting for an epoch to complete. 
\subsection{Revisiting Laplacian Eigenmaps and t-SNE}
Laplacian Eigenmaps is a graph based method with an emphasis on preserving the local structure of the manifold and encouraging the discovery of natural clusters in the dataset. Let $y_i$ be the $i^{th}$ sample from embedding $Y$, then the minimization of objective $J(y)$ (Eq:~\ref{eq:LE}) ensures that neighbouring samples $(\mathcal{A}_{ij} = 1)$ in the graph have encodings similar to each other. The choice of neighbours for a given sample are done by picking the top \textit{k} samples having the lowest euclidean distance to the fixed sample. The traditional solution involves constraint optimization which results in the eigendecomposition of the graph laplacian $(\mathcal{L})$.
\begin{equation}
\label{eq:LE}
\ensuremath{J(y) = \sum_{i, j}(y_i - y_j)^2\mathcal{A}_{ij} = 2Y^T\mathcal{L}Y}
\end{equation}
\begin{figure}
	\begin{minipage}{\linewidth}
		\includegraphics[scale=0.106]{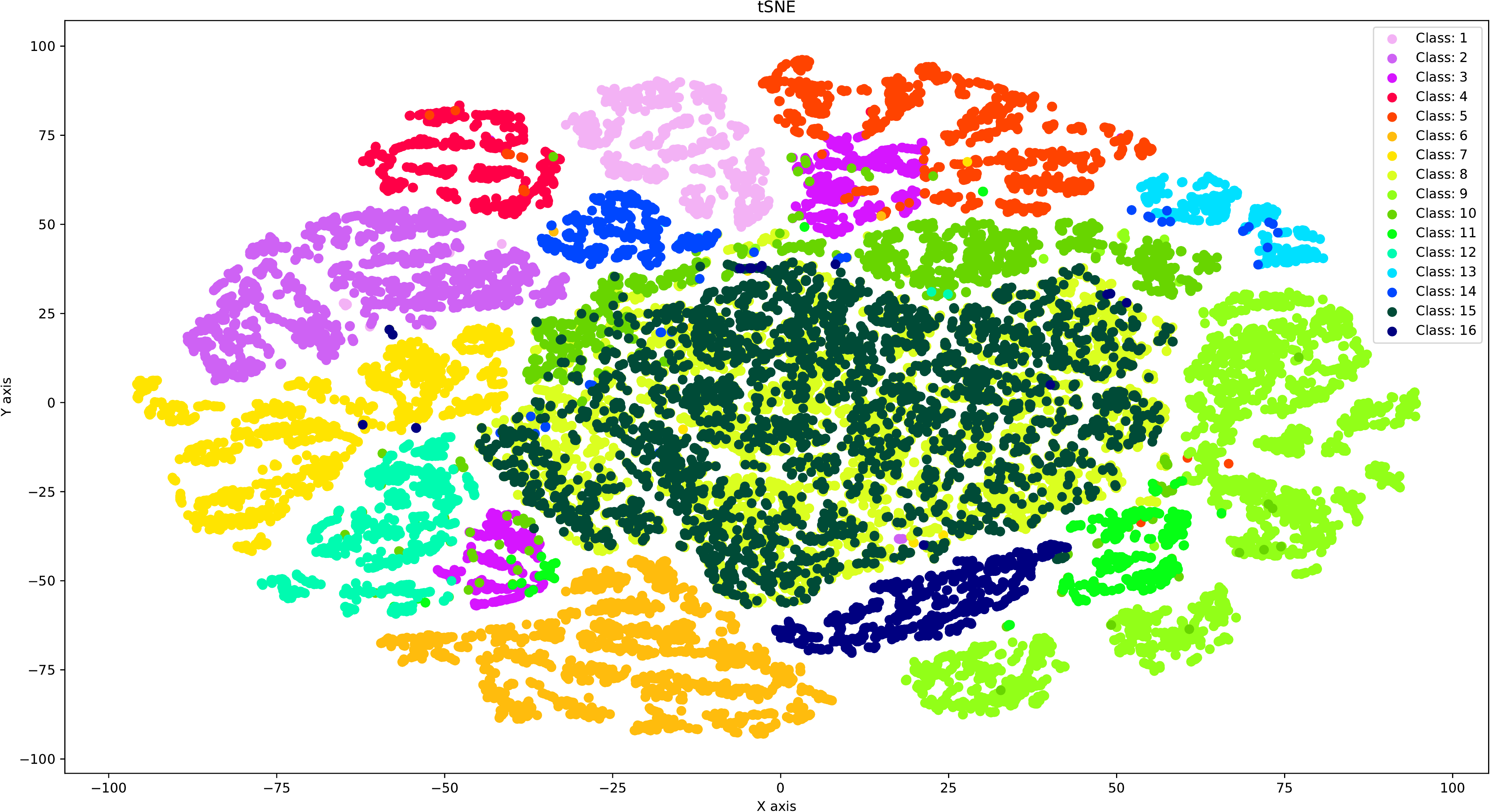}
		\includegraphics[scale=0.106]{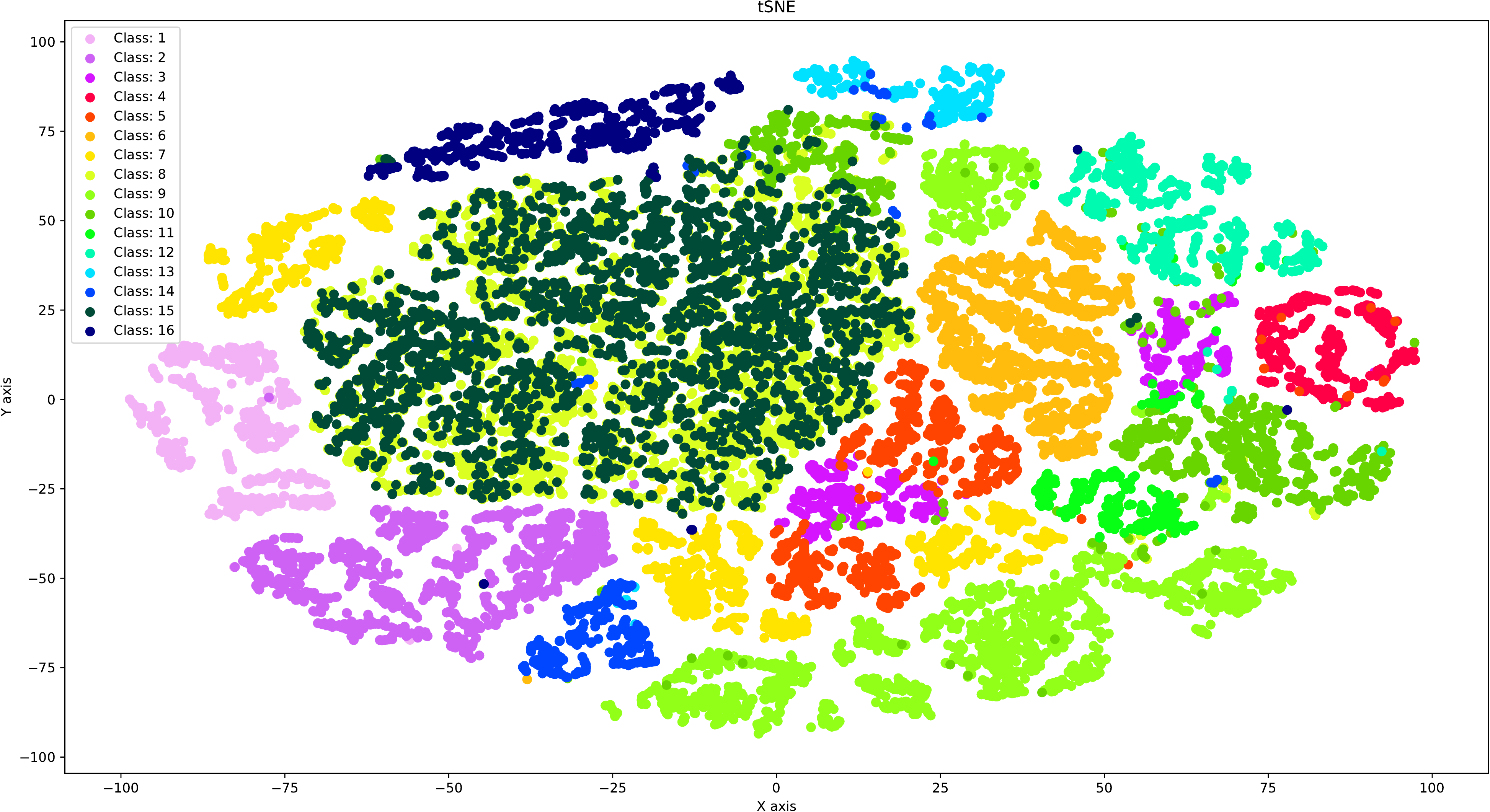}
	\end{minipage}
	\caption[center]{t-SNE visualization of Salinas: Notice the inconsistencies in the spatial relation between classes across multiple runs.}
	\vspace*{-5mm}
	\label{fig:tSNEcod}
\end{figure}
t-SNE is another popular algorithm for manifold visualization,which gives euclidean distances between samples a probabilistic interpretation. Let $x_i$ be the $i^{th}$ sample of $X$ in $\mathbb{R}^n$, then the probability of $x_j$ being a neighbour of $x_i$ is given in Eq:~\ref{eq:sne}. The variance $\sigma_i^2$ is a loose interpretation of the density of samples around $x_i$. A similar \textit{t-distribution} $q_{j|i}$ is computed for $y \in \mathbb{R}^m (m<n)$, followed by minimizing the KL divergence between $p_{ij}$ and $q_{ij}$. The resultant $y$ obtained are representative of the graph structure in $x$.
\begin{equation}
\label{eq:sne}
\ensuremath{p_{j|i} = \dfrac{exp(-||x_i - x_j||^2/2\sigma_i^2)}{\sum_{k \neq i}exp(-||x_i - x_k||^2/2\sigma_i^2)}}
\end{equation}
\subsection{LEt-SNE}
Our preliminary attempts at a parameterized algorithm began with Laplacian Eigenmaps. As before, Let $Y = f(X, w)$ then the objective (Eq:~\ref{eq:LE}) reduces to minimizing $\nabla_wY^TLY$ using gradient descent. In its current form, reducing $||w||$ will minimize the objective without the network learning anything meaningful. We corroborate this with experiments, which confirm that the network 'cheats' by either suppressing the L2-norm of the weights or collapsing all samples into the same region. Optimizing the network with respect to the constraints $Y^T\mathcal{D}Y=I$ \cite{BelkinNiyogi} ($\mathcal{D}$ is the Degree matrix) and $||w||_2 = k$ failed as it prevented the loss from converging to lower values. A key takeaway from this experiment was Laplacian Eigenmaps' ability to form tight cluster of embeddings for neighbouring samples in $X$. The question arises, can we devise a method on how to keep dissimilar points apart?
\par
We turn our attention to SNE and see how it alleviates this problem. If all samples $X$ are collapsed into very similar encodings $Y$, then the euclidean distance between an encoding $Y$ and all other points will not vary significantly. Thus, using Eq:~\ref{eq:sne}, $q_{j|i}\approx\, 1/|Y|\,\,\forall j$ i.e, the neighbourhood distribution for encodings $Y$ will approximate a uniform distribution. The objective of SNE hence imposes a large penalty on the collapse of embeddings into a small region. However, by introducing t-SNE in our solution, we also need to address the curse of dimensionality.  
\par
Recall the curse of dimensionality; as the dimensionality of our data increases, the variation in the inter-sample distance decreases. A similar scenario as the one discussed previously arises, this time among the samples in $X$. Therefore, $p_{j|i}\approx\, 1/|X|\,\,\forall j$, which leads to t-SNE giving inaccurate visualizations\footnote[2]{This holds true in a simplified scenario so as to provide intuition as to why t-SNE does not perform well in high-dimensional applications}. The effect of the curse of dimensionality leads to ambiguity in the relation between classes as shown in Fig:~\ref{fig:tSNEcod}. This leads us to the next question, could we stretch inter-sample distances to beat the curse of dimensionality?
\par
\begin{figure}[t!]
	\begin{minipage}{\linewidth}
		\centering
		\includegraphics[width=7.0cm]{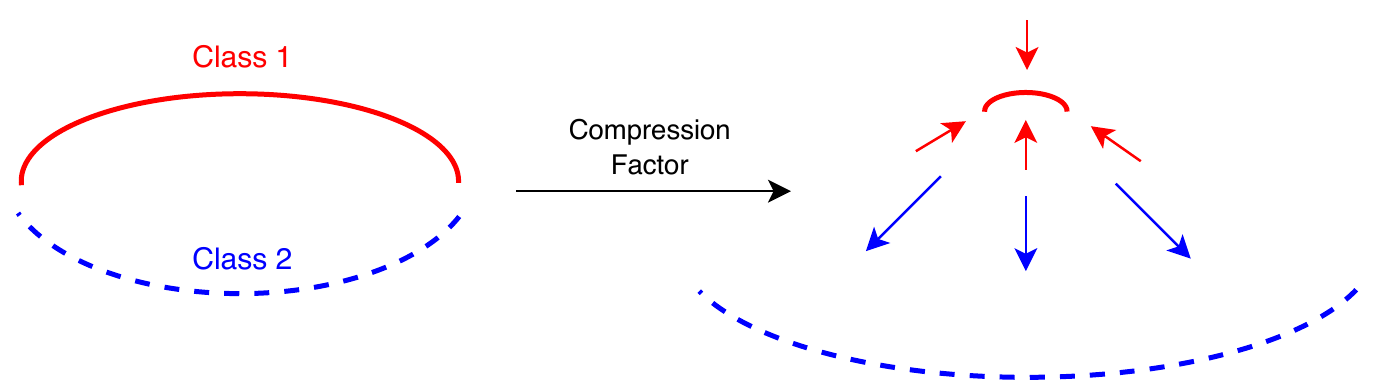}
	\end{minipage}
	\caption[center]{Compression acting on Class 1 samples. In practice, compression dominates as outward expansion from all the samples cancel out.}
	\vspace*{-5mm}
	\label{fig:CF}
\end{figure}
Our solution lies in defining a new term, \textit{Compression Factor}. The Compression Factor ($CF$) uses the Adjacency matrix $(\mathcal{A})$ generated by Laplacian Eigenmaps to give an illusion of manipulating the distance between samples in $\mathcal{X}$. We scale up the values of $p_{j|i}$ if $x_i$ and $x_j$ are connected in $\mathcal{A}$. This compression is mathematically defined in Eq:~\ref{eq:cf}. As a consequence we find that if $CF>1$; $\forall j \in neighbour\,(i)$, $\tilde{p}_{j|i} \uparrow$, whereas $\forall k \notin neighbour\,(i)$, $\tilde{p}_{k|i} \downarrow$. This approach helps in limiting the effect of curse of dimensionality, by creating the illusion that the difference in sample distances is larger than they appear, as shown in Fig: ~\ref{fig:CF}. 
\begin{equation}
\label{eq:cf}
\ensuremath{\tilde{p}_{j|i} = \frac{p_{j|i}*\{(CF- 1)*\mathcal{A}_{ij}+ 1\}}{\sum_j p_{j|i}*\{(CF- 1)*\mathcal{A}_{ij}+ 1\}}}
\end{equation} 
We also modify t-SNE to retain the conditional probabilities as proposed in SNE instead of the joint probabilities due to the strong gradients obtained in comparison to the latter\cite{vanDerMaaten2008}. Although we do not fix any particular network architecture, we recommend the use of Batch Normalization\cite{ioffe2015batch} for rapid convergence as well as adapting to the scale specified. In LEt-SNE, perplexity plays the role of determining the scale of our embeddings instead of translating to the number of neighbors as in t-SNE. In the next subsections, we describe the three modes of operation of the algorithm.
\subsubsection{LEt-SNE for Manifold Visualization} 
The algorithm for manifold visualization is straightforward and is shown in Eq:~\ref{eq:man_vis}, where $p_{i|j},\,q_{i|j}$ are computed using Eq:\{~\ref{eq:sne}, ~\ref{eq:cf}\}. The Adjacency matrix $A$ is computed using the \textit{top-k} nearest neighbours. For this task, it is more suitable to keep a low value for the number of neighbours hyperparameter, as well as a low value of compression factor (\raisebox{-0.25\baselineskip}{\textasciitilde}5), to prevent the probability values from saturating and retain some structure from the original encodings.
\begin{equation}
\label{eq:man_vis}
	\ensuremath{w^*=arg\min_w\mathbb{E}_x\left( \mathcal{Y}^T\mathcal{L}\mathcal{Y} + \lambda\sum_{i, j}\tilde{p}_{i|j}log\frac{\tilde{p}_{i|j}}{q_{i|j}}\right)}
\end{equation}
\subsubsection{LEt-SNE for Labelled Clustering}
Instead of computing the adjacency for $X$ using the \textit{top-k} neighbours approach, we directly use class labels. The new adjacency matrix is defined as:
\[
A_{ij} = \left\{
\begin{array}{@{}l@{\thinspace}l}
\text{1}  &: \text{class}_\text{i} == \text{class}_\text{j}\\
\text{0} &: otherwise \\
\end{array}
\right.
\]
A high compression factor ($>50$) acts upon the new adjacency matrix, saturating the probabilities and creating an illusion of tight clustering between intra-class samples and large separation between inter-class samples. In case the samples of a class come from a multimodal distribution, we can divide the samples into subclasses each of which captures a single mode of the distribution. The multimodality of a class can be observed in the manifold visualization technique described earlier. To ensure that the embeddings adequately represent this illusion, we compute $KL(q||p)$ instead of $KL(p||q)$. With this change, if $p_{j|i}$ is small (as is the case with inter-class separation), the corresponding $q_{j|i}$ too has to be a small value to prevent incurring a large loss. An explanation can be found in \cite{Goodfellow-et-al-2016}. The objective for minimization is: 
\begin{equation}
\label{eq:lab_clus}
\ensuremath{w^*=arg\min_w\mathbb{E}_x\left( Y^T\mathcal{L}Y+\lambda\sum_{(i, j)} q_{i|j} log \frac{q_{i|j}}{\tilde{p}_{i|j}} \right)}
\end{equation}
Note that we do not use classification gradients to allow the network to explore spatial relations within the dataset. 
\subsubsection{LEt-SNE for Unlabelled Clustering}
So far, we have seen two methods to compute the Adjacency matrix: \textit{top-k} and class labels. Are there any alternative approaches to design the Adjacency matrix such that Eq:~\ref{eq:lab_clus} can be used for unlabelled data?
\par
Let us assume a pixel in an image ($I$) to belong to a particular class, then it is highly likely that its 8-neighbours belong to the same class too. We then partition the image into disjoint regions, with pixels (samples) within a region considered connected components when computing the adjacency matrix. Formally, let $I$ be the Image composed of our samples $X$, segmented into regions $R$ such that $I = R_0 \cup R_1 \dots R_n$ and $R_i \cap R_n = \varnothing\,\,\forall i,j \in n; i \neq j$. The adjacency matrix is defined as:
\[
A_{ij} = \left\{
\begin{array}{@{}l@{\thinspace}l}
\text{1}  &: \text{R}_{\textit{x}_\text{i}} == \text{R}_{\textit{x}_\text{j}}\\
\text{0} &: otherwise \\
\end{array}
\right.
\]
As a proof of concept, we use two segmentation algorithms: Watershed \cite{Beucher1994} and SLIC \cite{Achanta:149300}. We prevent oversegmentation in SLIC by employing Region Adjacency Graph and Graph Cut algorithm.
\begin{figure*}
	\begin{minipage}{\linewidth}
		\centering
		\includegraphics[width=0.4\textwidth]{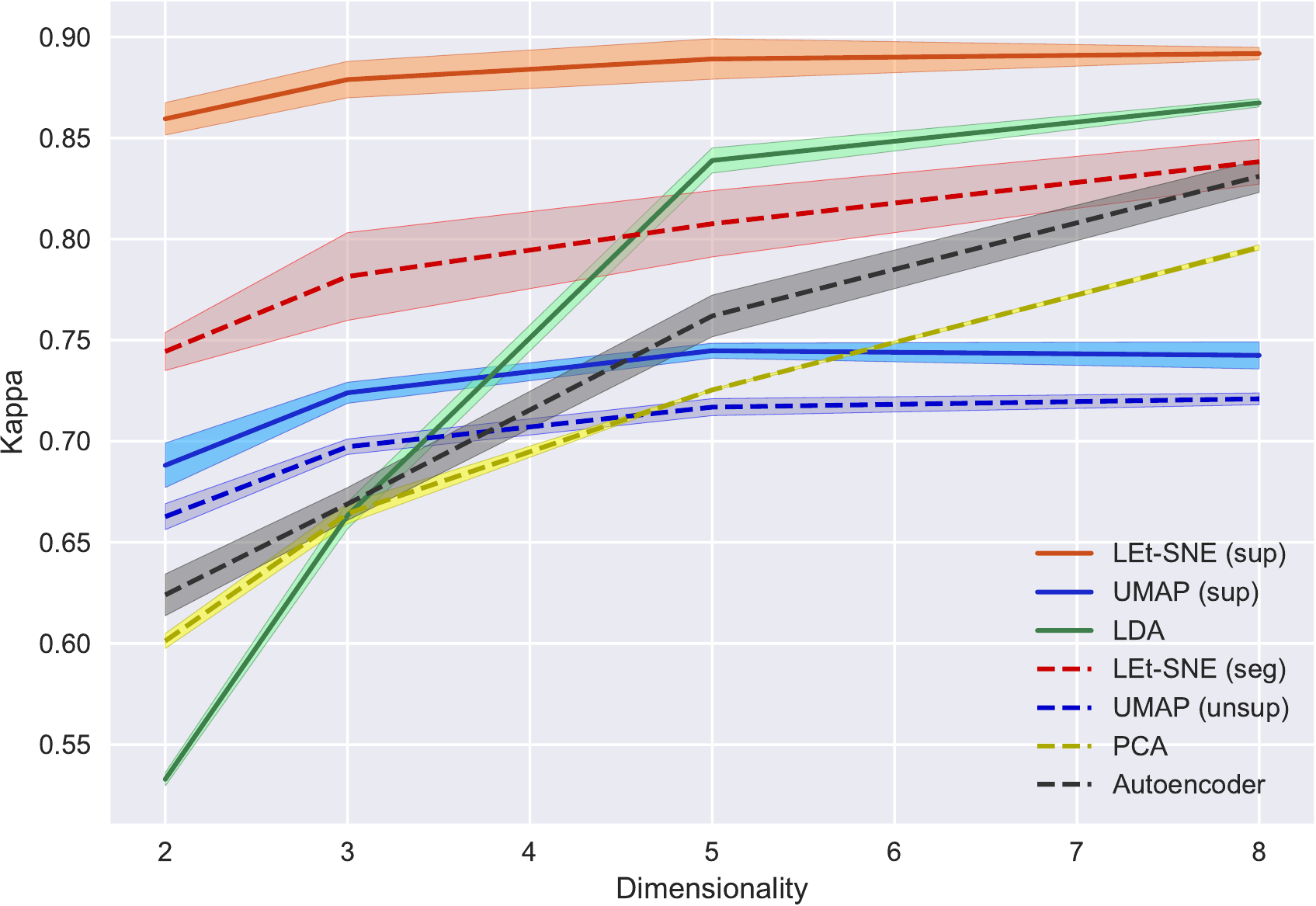}
		\includegraphics[width=0.4\textwidth]{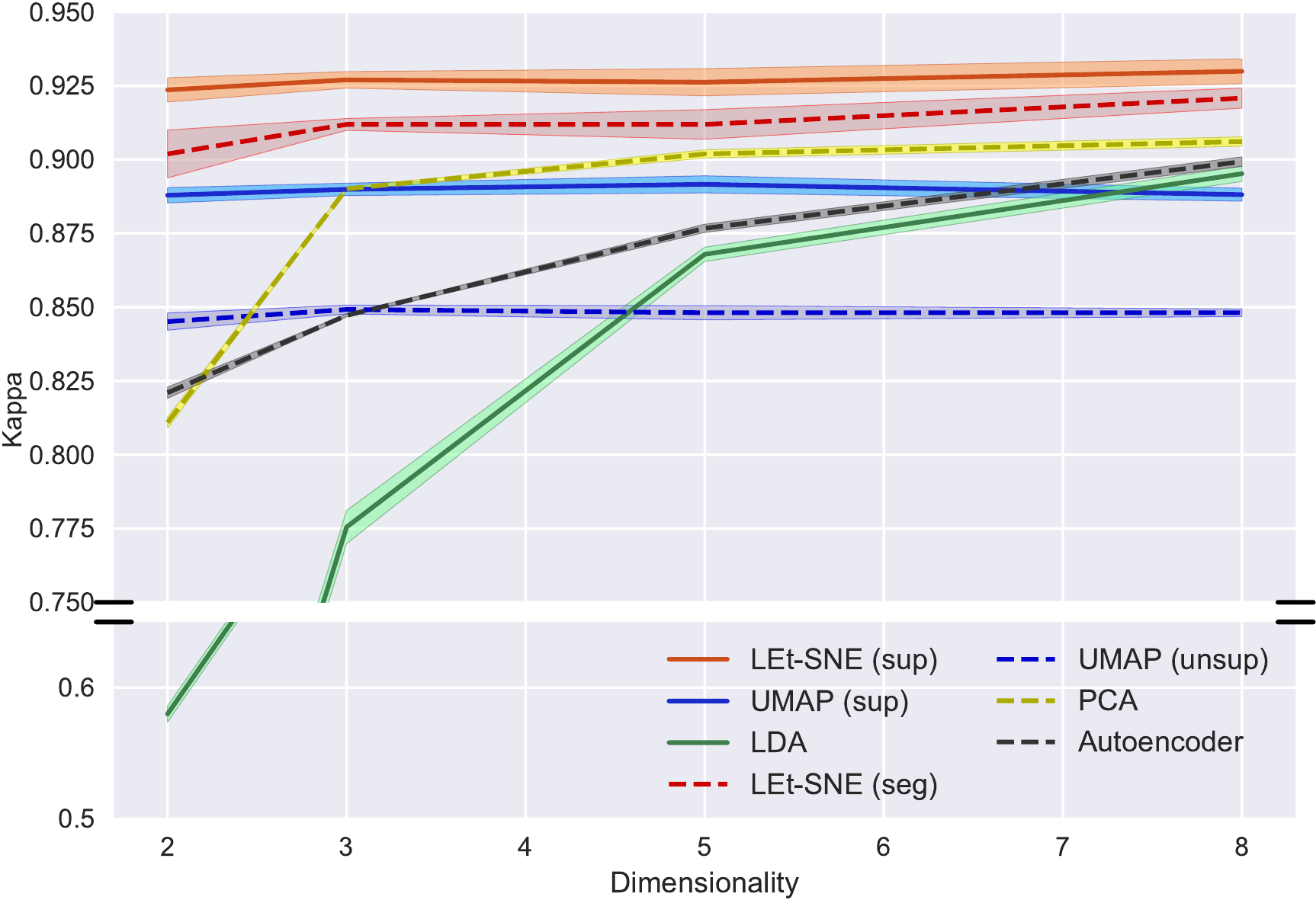}
	\end{minipage}
	\vspace*{-2mm}
	\caption[center]{Pavia (left) and Salinas (right): Comparing various supervised and unsupervised approaches}
	\vspace*{-5mm}
	\label{fig:kappa_plot}
\end{figure*}
\section{Experimentation and Results}
To keep the paper concise, we select a few experiments from each of the three modes of operation and produce them here. The code and the supporting material, including all experiments and class confusion maps\cite{MeghShuklaIITB} are available at \href{https://github.com/meghshukla/LEt_SNE}{GitHub: meghshukla/LEt-SNE}. 
\par
We use three datasets popular among the remote sensing community to verify our results: Indian Pines, Salinas and Pavia University. Indian Pines and Salinas features 16 classes each, with the former having considerably more overlap in classes than the latter. The Pavia University dataset contains 103 hyperspectral bands with 9 classes present. Further details on the datasets can be found in \cite{MeghShuklaIITB}. The data preprocessing step is limited to standardization with zero mean and unit variance. Our implementation is primarily based on TensorFlow. We use monte-carlo approximations of Eq:\{~\ref{eq:man_vis}, ~\ref{eq:lab_clus}\} over mini-batches $m$ for optimizing the weights $w$.
\par
\begin{figure}
	\begin{minipage}{\linewidth}
		\centering
		\begin{subfigure}{0.48\textwidth}
			\centering
			\includegraphics[width=\textwidth]{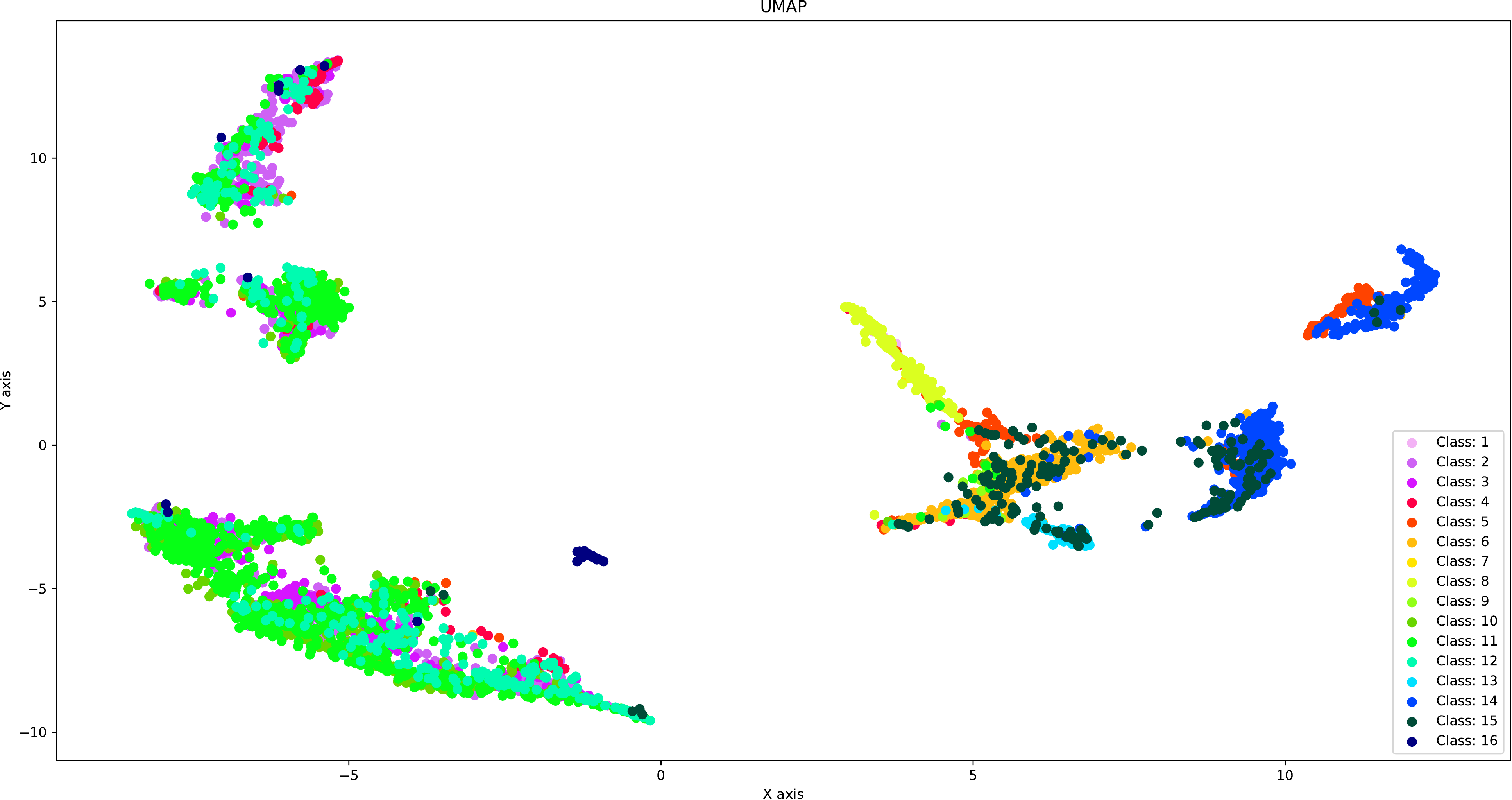}
			\caption{UMAP}
		\end{subfigure}
		\begin{subfigure}{0.48\textwidth}
			\centering
			\includegraphics[width=\textwidth]{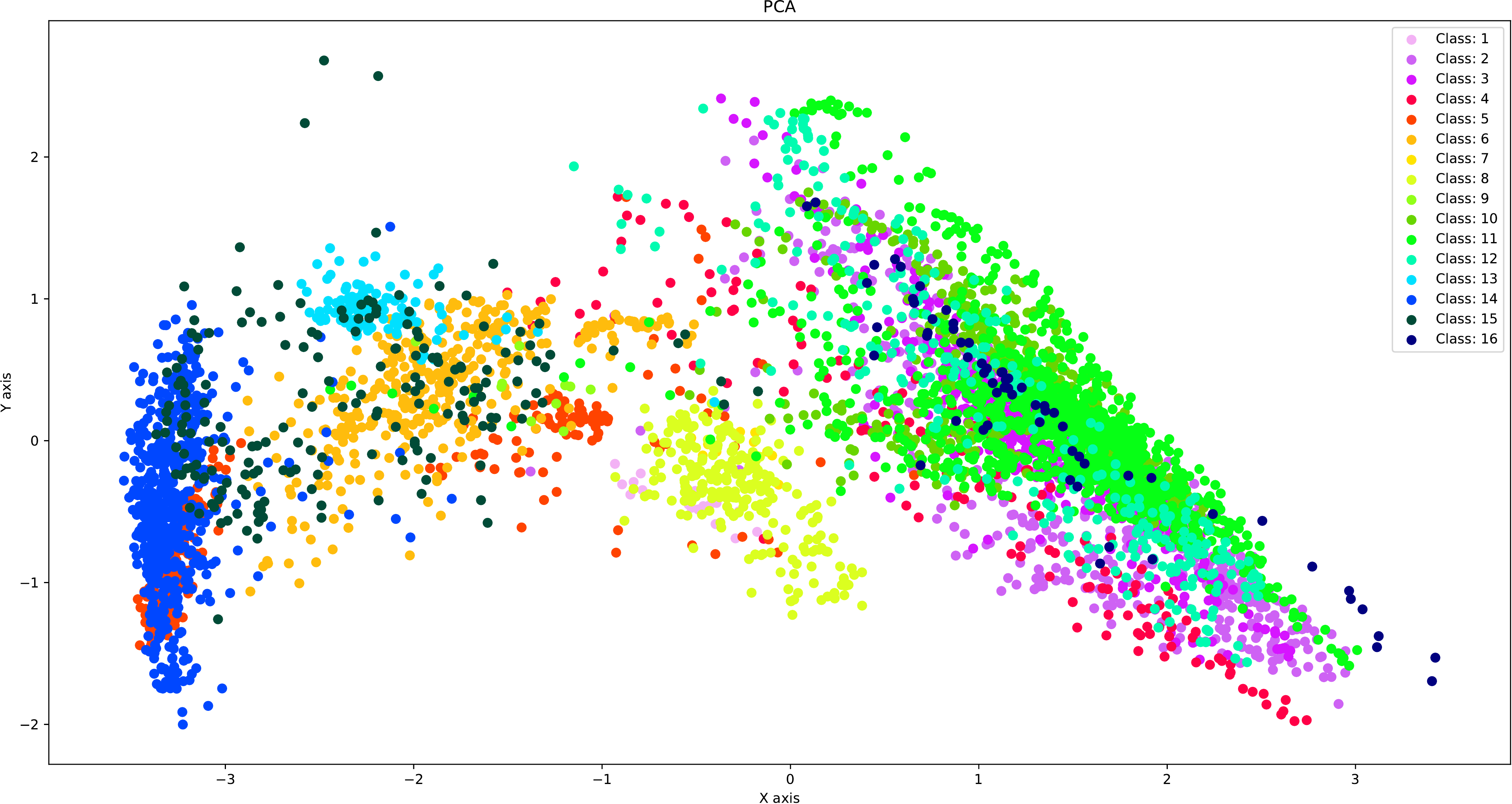}
			\caption{PCA}
		\end{subfigure}
		\begin{subfigure}{0.48\textwidth}
			\centering
			\includegraphics[width=\textwidth]{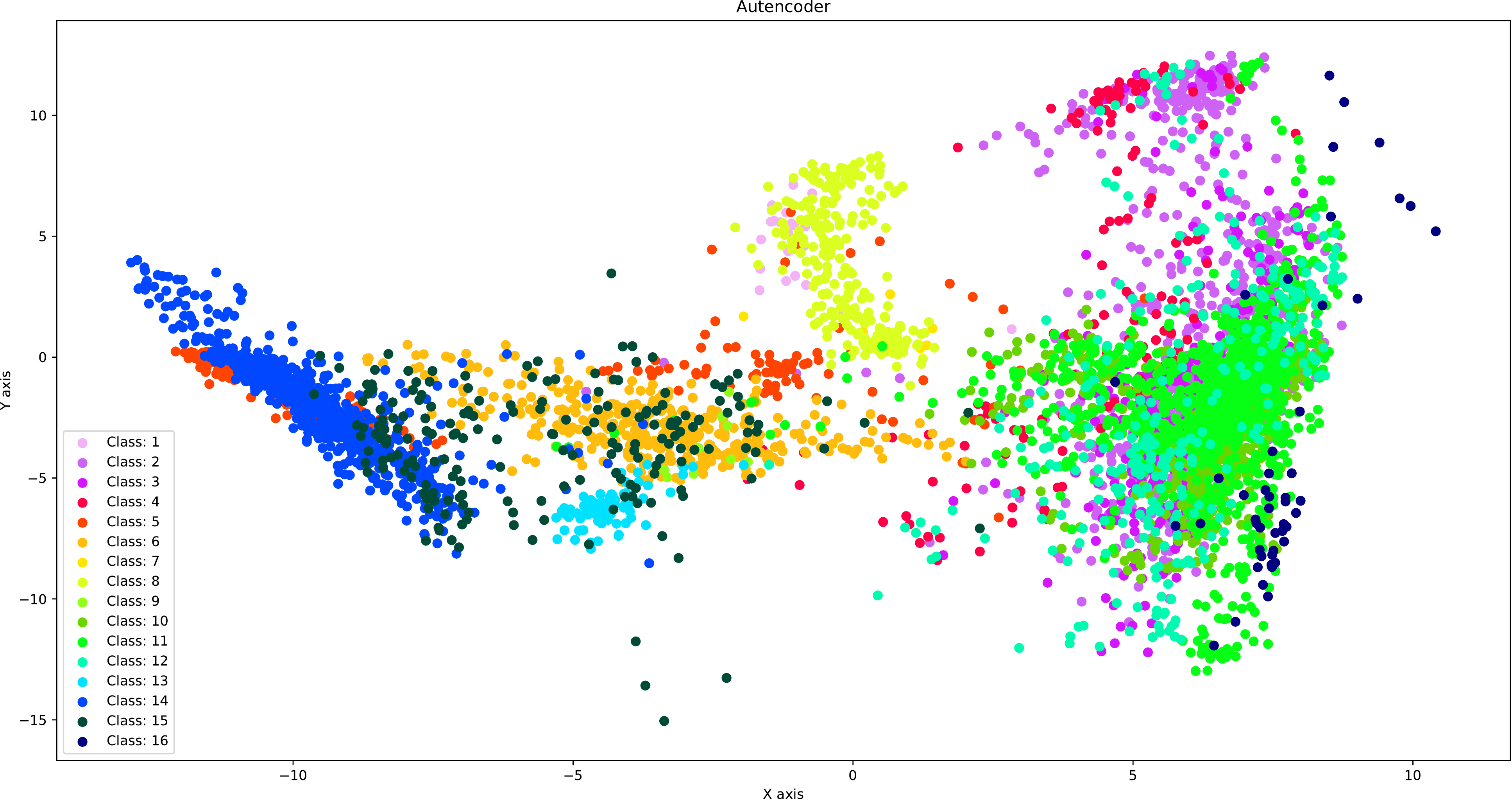}
			\caption{Autoencoder}
		\end{subfigure}
		\begin{subfigure}{0.48\textwidth}
			\centering
			\includegraphics[width=\textwidth]{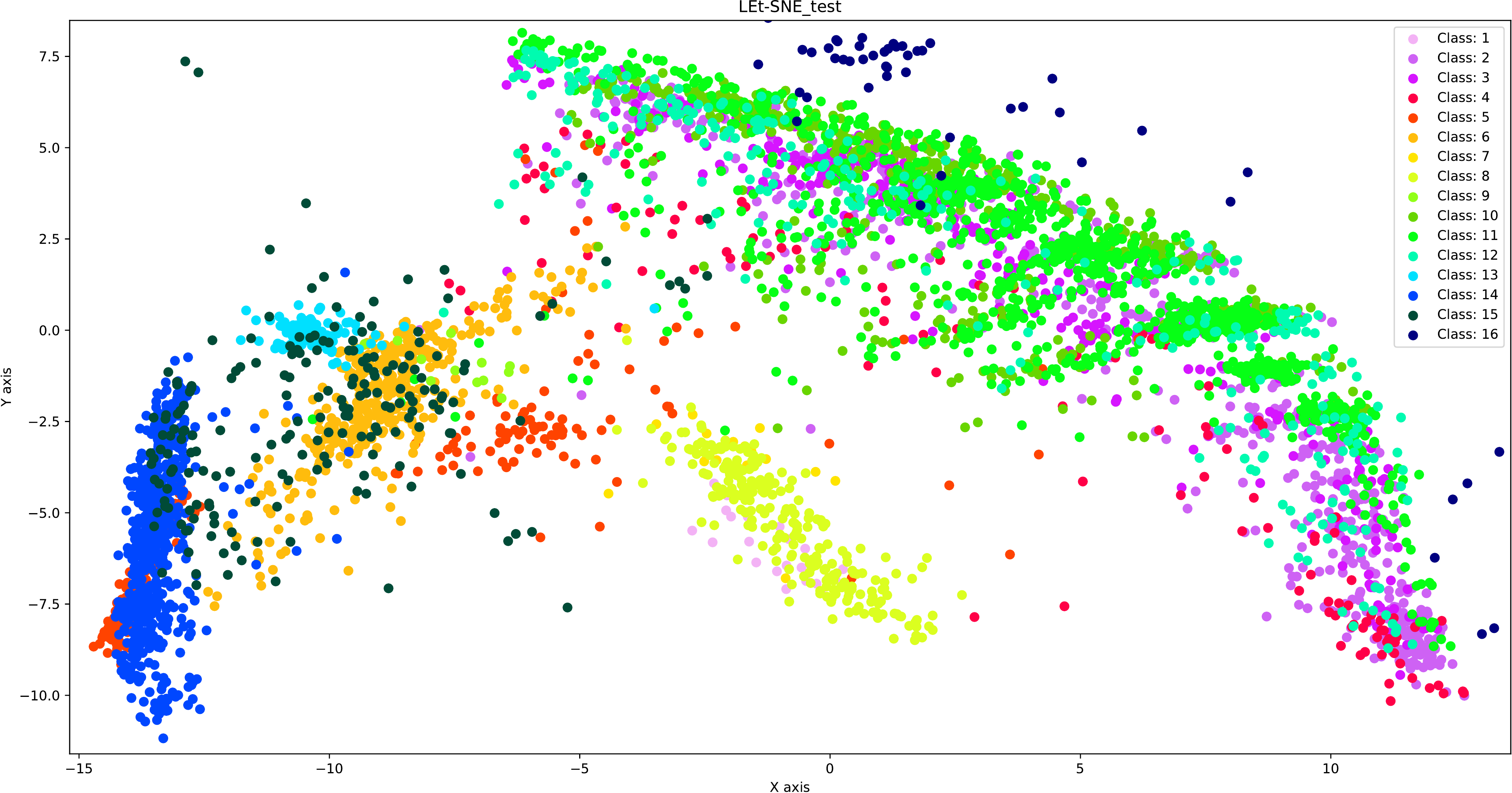}
			\caption{LEt-SNE}
		\end{subfigure}
	\end{minipage}
	\vspace*{-2mm}
	\caption{Indian Pines: Manifold Visualization}
	\vspace*{-2mm}
	\label{fig:man_vis_ip}
\end{figure}

\textbf{Manifold Visualization: }The two dimensional embeddings of Indian Pines are shown in Fig:~\ref{fig:man_vis_ip}. We use visual inspection to analyze the quality of embedding as done in \cite{vanDerMaaten2008}. We note that all approaches show similar characteristics, such as elongated strips of \textit{Classes: 10-12} on one side and \textit{Class 14}. UMAP visualization clusters the classes together, but lacks the fine structure as shown in LEt-SNE. On the other hand, LEt-SNE captures the local structure as well as global structure to a large extent. 
\par
\begin{figure}
	\begin{minipage}{\linewidth}
		\centering
		\begin{subfigure}{0.48\textwidth}
			\includegraphics[width=\textwidth]{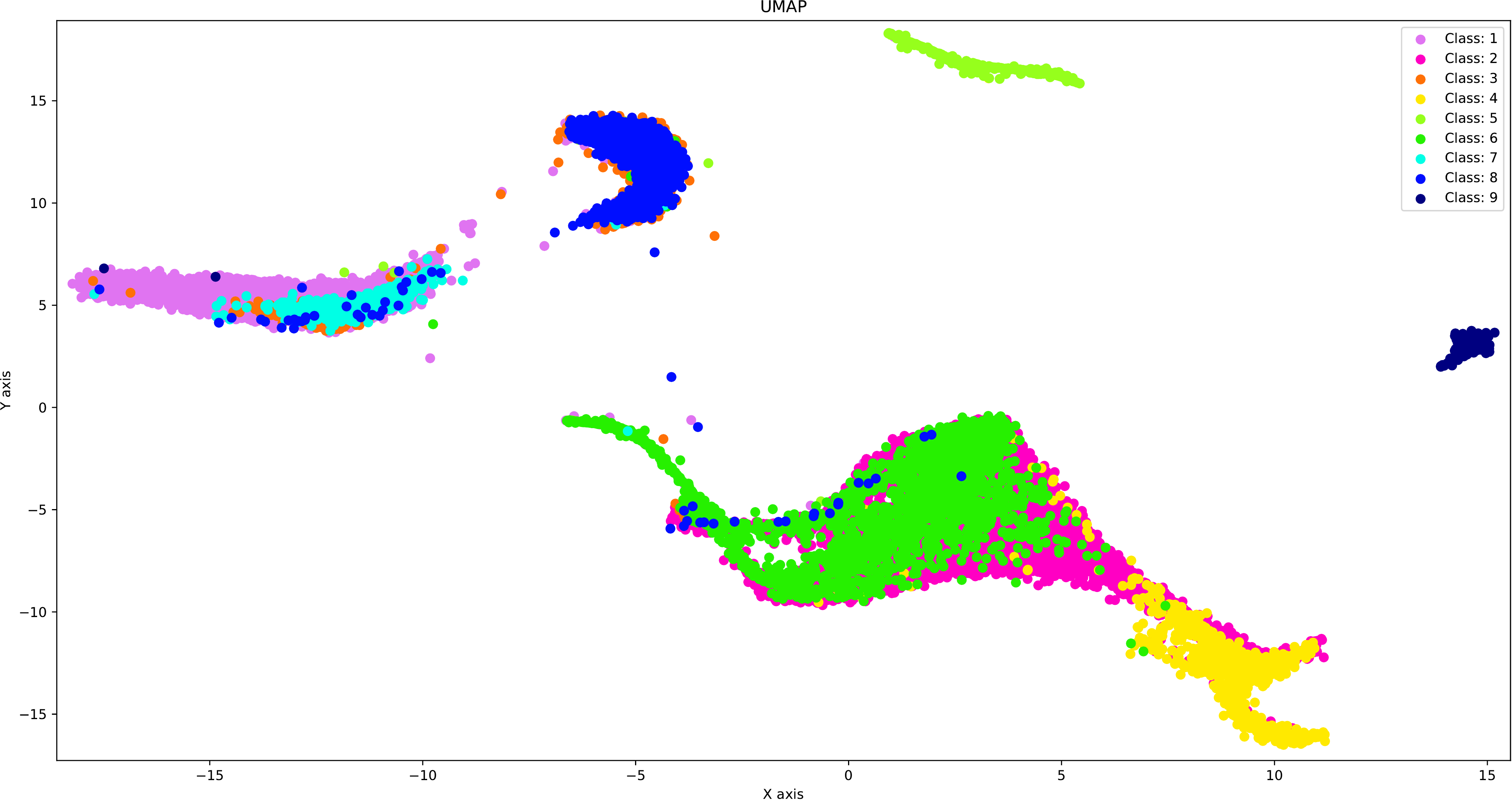}
			\caption[right]{UMAP}
		\end{subfigure}
		\begin{subfigure}{0.48\textwidth}
			\includegraphics[width=\linewidth]{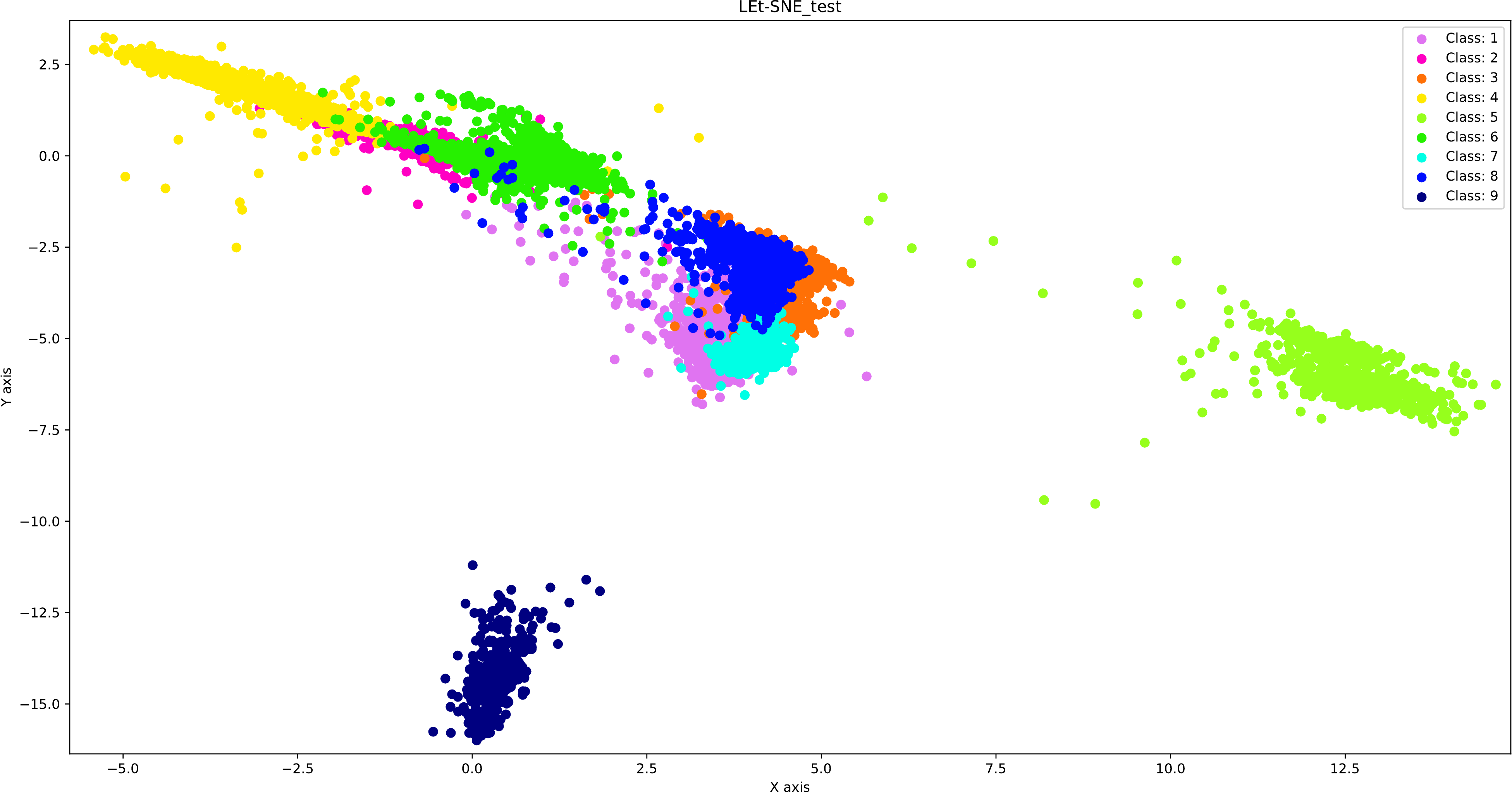}
			\caption[left]{LEt-SNE}
		\end{subfigure}
	\end{minipage}
	\vspace*{-2mm}
	\caption[center]{Pavia: Clustering with labels}
	\vspace*{-5mm}
	\label{fig:sup_pav}
\end{figure}

\textbf{Clustering with labels: } We evaluate the separation of classes and quality of clustering by training and evaluating the embeddings using a SVM classifier. The evaluation results for Pavia University and Salinas dataset are shown in Fig:~\ref{fig:kappa_plot}. We note that LEt-SNE (sup) outperforms all approaches, indicating better separation of samples using class labels. An intuition behind the results can be obtained by visualizing the embeddings shown in Fig:~\ref{fig:sup_pav}. We see that LEt-SNE provides better clustering and discriminative power between classes in comparison to UMAP, which is also verified in the confusion map. The importance of \textit{Compression Factor} is apparent from Table: ~\ref{tab:cf_tab}, where we note a significant improvement in performance of the algorithm.
\par
\begin{figure}
	\begin{minipage}{\linewidth}
		\begin{subfigure}{0.28\textwidth}
			\includegraphics[width=0.65\textwidth]{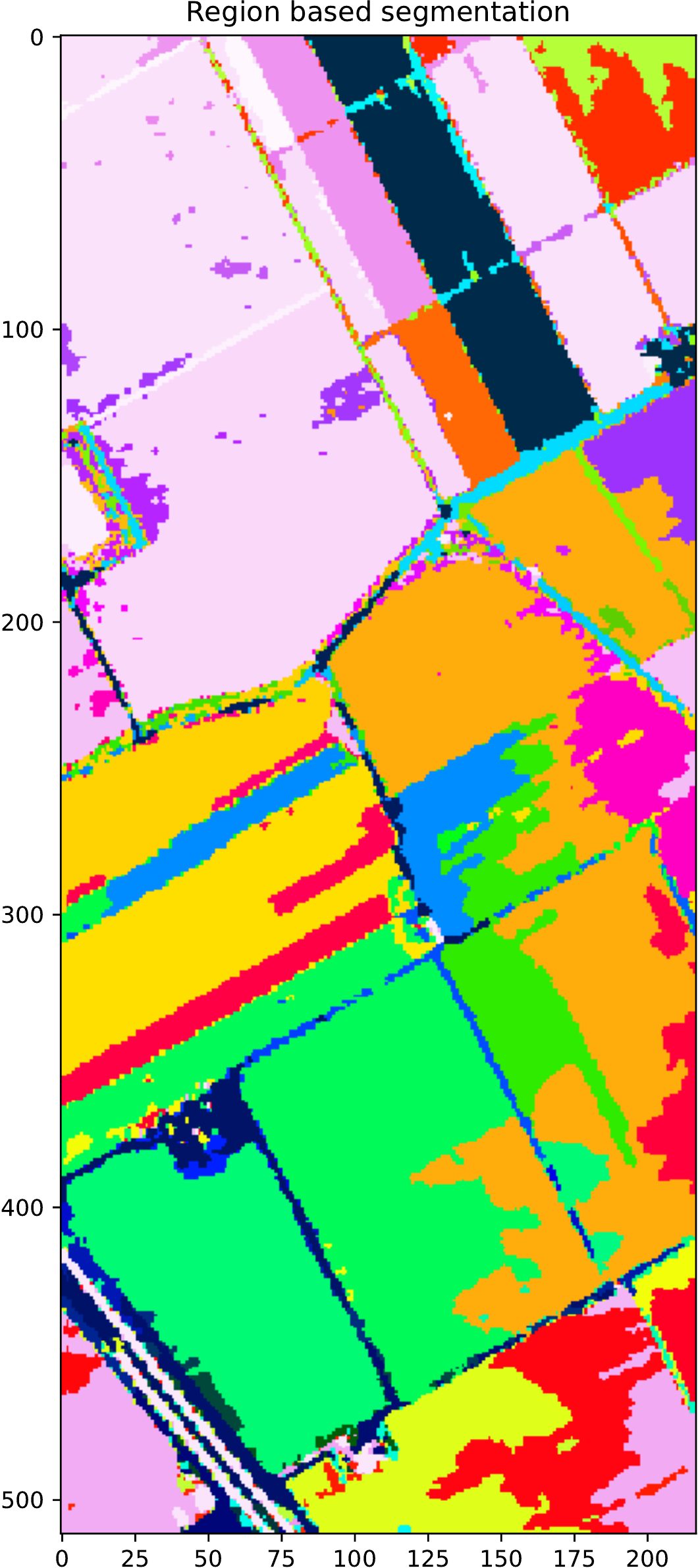}
		\end{subfigure}
		\begin{subfigure}{0.7\textwidth}
			\includegraphics[width=\textwidth]{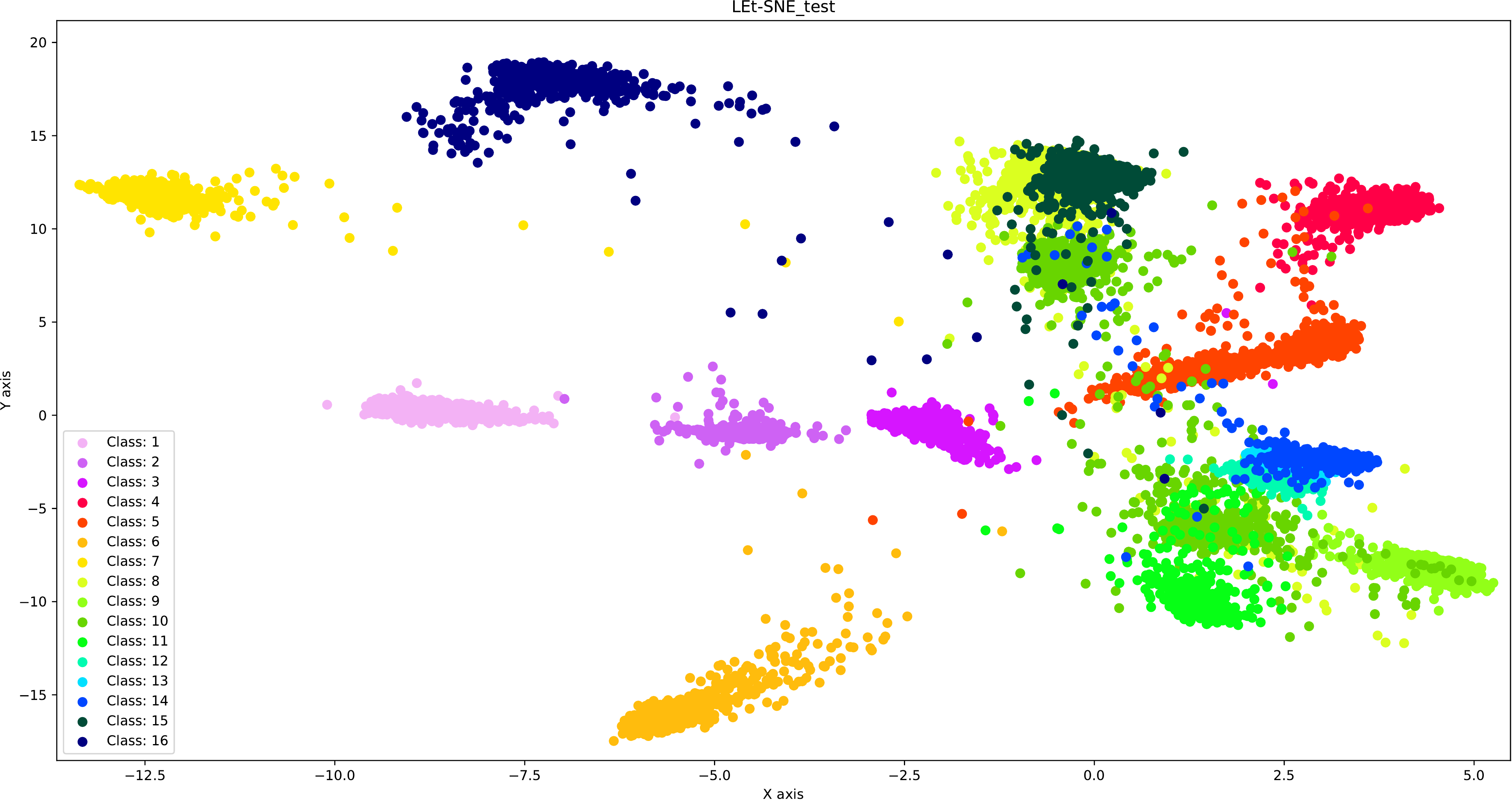}
		\end{subfigure}
	\end{minipage}
	\caption[center]{Salinas: (Left) Color coded disjoint regions\newline (Right) LEt-SNE embeddings}
	\vspace*{-5mm}
	\label{fig:seg_salinas}
\end{figure}

\textbf{Clustering without labels: }Depending on the choice of algorithm, a 3-channel (for SLIC) or 1-channel (for Watershed) image is provided as input for segmentation. The first principal component obtained from transforming the original channels using PCA, or the grayscale of the False Color Composite (FCC) image could be used as the 1-channel input. Similarly, the first three principal components or the FCC could be used as a 3-channel input based on which segmentation is performed. For Salinas, we use the grayscale image with Watershed algorithm, whereas for Pavia and Indian Pines we use the Princpal Components and SLIC for segmentation. The region segmentation and embeddings for Salinas dataset in shown in Fig:~\ref{fig:seg_salinas}. Refer to Fig:~\ref{fig:kappa_plot}, where we notice that even in the absence of labels, the segmentation based adjacency matrix used by LEt-SNE (seg) provides vital information which is used by compression factor to provide meaningful embeddings.

\section{Conclusion}
In this work, we have attempted to solve the problem of dimensionality reduction by proposing a new method, LEt-SNE. We have focused on parameterization, computational time, curse of dimensionality and producing intuitive embeddings when designing the algorithm. We have successfully demonstrated the use of Compression Factor to help alleviate the curse of dimensionality. With LEt-SNE, we solve a three-fold problem: Manifold Visualization, Clustering with Labels and Clustering without labels; thereby extending on the use cases of t-SNE. Our results show that LEt-SNE is competitive with popular state-of-the-art algorithms on common remote sensing datasets.
\begin{table}
	\centering
	\caption{Accuracy and Compression Factor: LEt-SNE (sup) with Dimensions = 2}
	\vspace*{-2mm}
	\label{tab:cf_tab}
	\footnotesize
	\begin{tabular}{cccc}
		\toprule 
		Compression & Indian Pines & Salinas & Pavia\\
		\midrule
		\textbf{NA} & 0.4936 & 0.7877 & 0.7534\\
		\textbf{200} & \textbf{0.6207} & \textbf{0.9236} & \textbf{0.8594}\\
		\bottomrule
		\vspace*{-5mm}
	\end{tabular}
\end{table}
\newpage  
\bibliographystyle{IEEEbib}
\bibliography{strings,refs}
\end{document}